\documentclass[%
 reprint,
 amsmath,amssymb,
 aps,
 prl,
nolongbibliography
]{revtex4-2}

\usepackage{graphicx}
\usepackage{dcolumn}
\usepackage{bm}

\usepackage{color}

\newcommand{\nc}{\newcommand}
\nc{\bra}{\langle}
\nc{\ket}{\rangle}
\nc{\vac}{|0\ket}
\nc{\da}{^{\dagger}}
\nc{\ps}{\hat{\psi}}
\nc{\pd}{\hat{\psi}\da}
\nc{\HLL}{\hat{\mathcal{H}}_{\text{LL}}}
\nc{\red}{\textcolor{red}}
\nc{\blue}{\textcolor{blue}}
\nc{\sred}[1]{\textcolor{red}{\sout{#1}}}
\nc{\sblue}[1]{\textcolor{blue}{\sout{#1}}}
\nc{\Imath}{\mathcal{I}}

\begin{document}

\title{Multi-Quantum Dark Solitons in One-Dimensional Bose Gas}

\author{Yuki Ishiguro$^1$}
\author{Jun Sato$^2$}
\author{Katsuhiro Nishinari$^3$}
 \affiliation{$^1$Department of Aeronautics and Astronautics, Faculty of Engineering, The University of Tokyo, 7-3-1 Hongo, Bunkyo-ku, Tokyo 113-8656, Japan\\
 $^2$Faculty of Engineering, Tokyo Polytechnic University, 1583 Iiyama, Atsugi, Kanagawa 243-0297, Japan\\
 $^3$Research Center for Advanced Science and Technology, The University of Tokyo, 4-6-1 Komaba, Meguro-ku, Tokyo 153-8904, Japan
 }

\begin{abstract}
Quantum and classical integrable systems share common mathematical structures, and the phenomena appearing in them are interrelated.
Solitons, which universally appear in classical integrable systems, also appear in quantum integrable systems.
Here, we consider quantum-classical correspondence in a one-dimensional Bose gas with repulsive delta-function interaction and present quantum states corresponding to multi-dark solitons.
Using an exact method, we compute the time evolution of the density profile in the multi-quantum dark soliton states. 
Localized solitary waves that behave like classical dark solitons are observed in the density profile.
We observe collisions of quantum dark solitons and show that they exhibit the properties of classical solitons: stability against scatterings and position shifts due to interactions.
\end{abstract}

\maketitle


\textit{Introduction.}---
Integrable systems contain an enormous number of conserved charges and exhibit interesting atypical phenomena.
Solitons are examples of such phenomena that appear universally in classical integrable systems \cite{faddeev2007hamiltonian,miwa2000solitons}.
The meaning of integrability differs between classical and quantum integrable systems.
Classical integrable systems are differential equations for which exact solutions can be constructed.
By contrast, quantum integrable systems are systems in which the Hamiltonian is exactly diagonalizable.
Although the problem settings of quantum and classical integrable systems are quite different, the physical phenomena that appear in them are closely related.
Solitons, which are exact solutions of classical integrable equations, have also been observed in quantum integrable systems theoretically \cite{wadachi1984classical,wadati1985quantum,PhysRevA.40.854,Ishikawa,PhysRevLett.108.110401,Sato_2016,PhysRevA.99.043632,PhysRevResearch.2.033368,Kaminishi_2020,universe8010002,PhysRevA.79.063616,PhysRevA.81.023625,PhysRevA.92.032110,PhysRevA.94.023623,ayet2017single,PhysRevB.92.214427} and experimentally \cite{drummond1993quantum,nguyen2014collisions,PhysRevLett.83.5198,PhysRevLett.101.130401,becker2008oscillations,Frantzeskakis_2010}.
Revealing how solitons emerge in quantum integrable systems is an important problem to achieve a deep understanding of the physics in integrable systems.

The Lieb--Liniger (LL) model, which describes bosons interacting with a delta-function potential in one dimension, is one of the most fundamental quantum integrable systems in the investigation of quantum-classical correspondence of solitons \cite{PhysRev.130.1605,PhysRev.130.1616}. 
The time evolution equation of the field operator in the LL model corresponds to the nonlinear Schr\"{o}dinger (NS) equation \cite{1972JETP3462Z,tsuzuki1971nonlinear}.
The NS equation is a classical integrable system and includes two types of solitons: bright solitons \cite{1972JETP3462Z} and dark solitons \cite{tsuzuki1971nonlinear}.
The problem of constructing quantum states corresponding to these solitons has been discussed over several decades \cite{wadachi1984classical,wadati1985quantum,PhysRevA.40.854,Ishikawa,PhysRevLett.108.110401,Sato_2016,PhysRevA.99.043632,PhysRevResearch.2.033368,Kaminishi_2020,universe8010002}.
The LL model is exactly solvable, and eigenstates of the Hamiltonian are obtained using the Bethe ansatz.
Since eigenstates are translationally invariant under the periodic boundary condition, the density profile of each eigenstate is flat.
However, it is known that quantum soliton states, in which the density profiles correspond to classical solitons, can be constructed by superposing certain eigenstates.
When the interaction is attractive, the LL model has string-type Bethe eigenstates that correspond to the bound states.
Quantum bright soliton states are constructed from superpositions of these eigenstates \cite{wadachi1984classical,wadati1985quantum,PhysRevA.40.854}.
By contrast, quantum dark soliton states are constructed from eigenstates of the LL model with repulsive interaction.
Single-quantum dark soliton states are constructed by superposing Bethe eigenstates of one-hole excitations called type II excitations \cite{PhysRevLett.108.110401,Sato_2016} by using the fact that dispersion relations of dark solitons in the NS equation correspond to those of type II excitations in the LL model \cite{Ishikawa}.
However, the construction of quantum states corresponding to multi-dark solitons has not yet been established, mainly because eigenstates corresponding to multi-dark solitons are not known. 
Although the construction of 2-quantum soliton states by superposing eigenstates of two-hole excitations was proposed in \cite{universe8010002}, the scattering of quantum solitons has not yet been realized with this approach.

One of the most outstanding features that characterizes solitons is their stability against scatterings. 
Localized solitary waves pass through each other without changing their shapes and velocities. 
The effect of interaction among solitons manifests itself in the position shifts.
Remarkably, this $N$-body scattering can be factorized as products of two-body interactions, which is the key constituent of the integrability. 
This factorization property is also essential for the quantum case. 
The position shift in the classical soliton is described through the two-body scattering $S$-matrix, which also plays a crucial role in the Bethe ansatz formulation in quantum integrable systems. 
It is an urgent task, therefore, to observe scatterings and position shifts in the quantum soliton states from the point of view of classical-quantum correspondence in the integrable systems. 

In this Letter, we present the construction of multi-quantum dark solitons and succeed in observing their scatterings in a quantum integrable system. 
By considering type II excitations as units and combining them, we introduce a new class of excitations corresponding to multi-dark solitons.
Moreover, we also present an approach to control the velocities of each solitary wave through the ``shift parameter" $\Delta$, which will be explained later. 
These enable us to realize scatterings of quantum dark solitons, and find that quantum dark solitons are stable against collisions and their positions are shifted due to interactions.
These observations strongly suggest that the proposed quantum states indeed correspond to the classical solitons.

\textit{Model.}---
We consider the LL model for $N$ bosons with repulsive interaction ($c>0$):
\begin{align}
    \HLL=-\sum_{j=1}^{N}\frac{\partial^2}{\partial x^2_j}+2c\sum_{j<k}^N \delta(x_j-x_k).
    \label{eq:Hamiltonian_1st}
\end{align}
Here, we impose the periodic boundary condition with the system size $L$ and use units with $2m=\hbar=1$ where $m$ is the mass of the particles. 
By introducing the Bose field operator $\ps(x,t)$, the Hamiltonian in the second quantization formalism is given by
\begin{align}
    \HLL=\int_0^L dx \left[ \partial_x \pd \partial_x \ps + c \pd \pd \ps \ps \right].
    \label{eq:Hamiltonian_2nd}
\end{align}
The Heisenberg equation of motion for the Bose field operator $\ps(x,t)$ is 
\begin{align}
    i\partial_t \ps = -\partial^2_x \ps + 2c\pd\ps\ps,
    \label{eq:NLS}
\end{align}
which corresponds to the NS equation.

The LL model is exactly solvable by the Bethe ansatz \cite{korepin_bogoliubov_izergin_1993}.
Each eigenstate of the Hamiltonian is determined by a set of Bethe quantum numbers $\{I\}_N=\{I_1,I_2,\cdots,I_N \}$. Here, $I_j$'s are integers (half-integers) for odd (even) particle number $N$. 
For a given set of $\{I\}_N$ chosen as $I_1<I_2<\cdots<I_N$, the Bethe equations: 
\begin{align}
    k_j L=2\pi I_j - \sum_{l \neq j}^N \text{arctan}\left( \frac{k_j-k_l}{c} \right),
    \label{eq:BAE}
\end{align}
have a unique real solution $\{k\}_N=\{k_1,k_2,\cdots,k_N \}$ such that $k_1<k_2<\cdots<k_N$.
The total momentum $P$ and the energy eigenvalue $E$ are expressed in terms of the Bethe roots $\{k\}_N$ as 
\begin{align}
    P=\sum_{j=1}^N k_j=\frac{2\pi}{L}\sum_{j=1}^N I_j,\quad E=\sum_{j=1}^N k_j^2.
\end{align}
The set of Bethe quantum numbers corresponding to the ground state is given by 
\begin{align}
\{I\}_{N,0}:=\left\{-\frac{N-1}{2}+j\middle| 0\le j\le N-1\right\}.   
\end{align}
Type II excitations are defined as one-hole excitations, the Bethe quantum numbers of which are given by
\begin{align}
\{I\}_{N,h}:=\left\{-\frac{N-1}{2}+j \middle| 0\le j\le N\right\} \setminus \left\{ \frac{N+1}{2} -h\right\},
\label{eq:typeII}
\end{align}
where $h$ ($0\le h\le N-1$) denotes the position of a hole.
The total momentum of $\{I\}_{N,h}$ is $P_h=\frac{2\pi h}{L}$.

\textit{Single-quantum dark soliton.}---
Single-quantum dark soliton states are constructed by superposing eigenstates of type II excitations.
Before introducing multi-quantum dark soliton states, we first extend the construction of single-quantum dark soliton states presented in \cite{PhysRevLett.108.110401,Sato_2016} for later discussion.

\begin{figure}[bt]
    \centering
    \includegraphics[height=9.0cm]{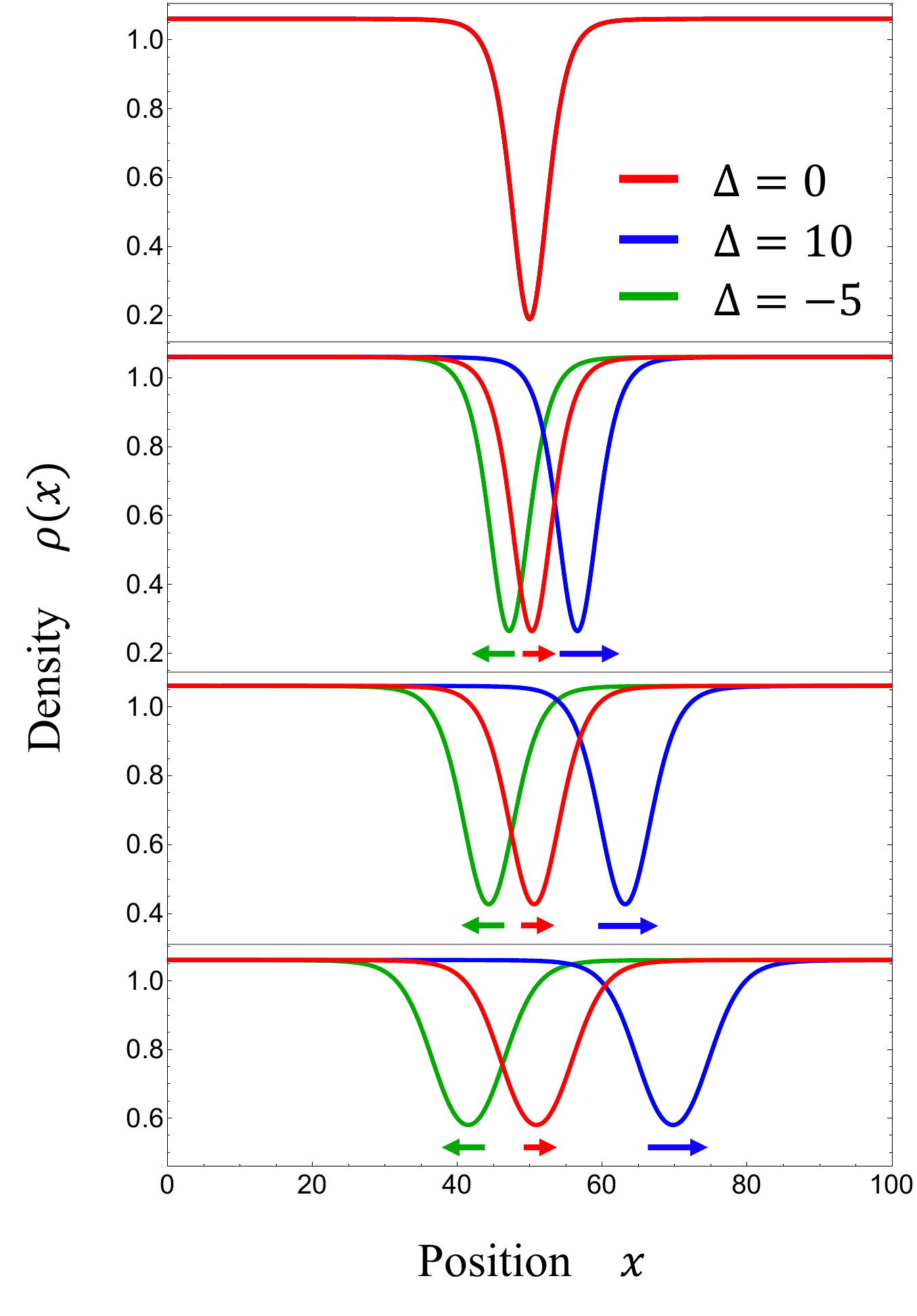}
    \caption{Time evolution of the density profile in the single-quantum soliton states. Parameters are set to $L=N=100$, $X=\frac{L}{2}$, and $c=0.1$.
    The shift parameters are $\Delta=0$ (red line), $10$ (blue line), and $-5$ (green line).
    While the initial density profiles are identical regardless of the value of $\Delta$, the velocities of the solitons depend on it.}
    \label{fig:Single-soliton}
\end{figure}

By introducing the shift parameter $\Delta (\in \mathbb{Z})$, which enables us to control the velocity of the soliton, we define a $\Delta$-shifted type II excitation with the total momentum $P_h=\frac{2\pi (h+N\Delta)}{L}$ as
\begin{align}
    \{I\}_{N,h,\Delta}:=\{I\}_{N,h}+\Delta,
    \label{eq:delta-typeII}
\end{align}
where $h$ ($0 \le h \le N-1$) indicates the hole position.
We define a family of sets of $\Delta$-shifted type II excitations with particle number $N$ as
\begin{align}
    \Imath_{N,\Delta}:=\{ \{I\}_{N,h,\Delta} | 0 \le h \le N-1 \}.
\end{align}
A single-quantum soliton state is constructed by superposing all eigenstates in $\Imath_{N,\Delta}$.
We denote the normalized Bethe eigenstate of the $\Delta$-shifted type II excitation $\{I\}_{N,h,\Delta}$ with the total momentum $P_h=\frac{2\pi (h+N\Delta)}{L}$ as $|P_h\ket$.
Then, we introduce a single-quantum soliton state $|X\ket$ as follows:
\begin{align}
    |X\ket=\frac{1}{\sqrt{N}} \sum_{h=0}^{N-1} \text{exp}(-i P_h Q)|P_h\ket,
    \label{eq:single-quantum-soliton}
\end{align}
where $Q$ is a parameter that determines the position of soliton $X$ as $X=Q+\frac{L}{2}$.
This is an extension of the conventional single-quantum dark soliton states proposed in \cite{PhysRevLett.108.110401,Sato_2016} with one parameter $\Delta$. 
When $\Delta=0$, (\ref{eq:single-quantum-soliton}) corresponds to the conventional case.
In the case of the conventional quantum soliton state, the expectation value of the density operator $\hat{\rho}(x)=\pd(x)\ps(x)$ in the initial state is known to correspond to a single-classical dark soliton in the weak coupling limit $c\to0$. 
Although a wave packet disperses due to quantum effects, the dynamics of a quantum soliton is also similar to that of a classical soliton.
The extended quantum soliton state (\ref{eq:single-quantum-soliton}) retains these properties.
Fig. \ref{fig:Single-soliton} shows the time evolution of the density profile in the single-quantum soliton states (\ref{eq:single-quantum-soliton}) with the shift parameters $\Delta=0$, $10$, and $-5$. 
The initial density profiles are identical regardless of the value of $\Delta$.
However, the soliton velocity depends on the value of $\Delta$. It can be observed in Fig. \ref{fig:Single-soliton} that the velocity of the soliton increases with the value of $\Delta$. 
Thus, the velocity of the quantum soliton can be controlled by adjusting the shift parameter $\Delta$.

\textit{Multi-quantum dark soliton.}---
Let us introduce the construction of multi-quantum soliton states.
As shown in Eq. (\ref{eq:single-quantum-soliton}), a single-quantum soliton state is constructed by superposing all eigenstates in $\Imath_{N,\Delta}$.
In the following, we will define a ``connected type II excitation'' that is created by a union of $n$ sets of type II excitations $\{I\}_{N_j,h_j,\Delta_j} \in \Imath_{N_j,\Delta_j}$ ($1\le j\le n$).
An $n$-quantum soliton state is constructed by superposing all eigenstates corresponding to connected type II excitations.

First, we divide $N$ particles into $n$ parts as $N=\sum_{j=1}^n N_j$. 
For simplicity, we assume that particle numbers $N$ and $N_j$ ($1\le j\le n$) are all even.
Then, we consider $\Delta$-shifted type II excitations $\Imath_{N_j,\Delta_j}$ ($1\le j\le n$).
Here, we need to select $\Delta_j$ such that the Bethe quantum numbers do not overlap: $\{I\}_{N_j,h_j,\Delta_j} \cap \{I\}_{N_k,h_k,\Delta_k} =\phi$ for $1\le \forall j< \forall k \le n$ and $0\le \forall h_j(\forall h_k) \le N_j-1(N_k-1)$, where $\{I\}_{N_j,h_j,\Delta_j} \in \Imath_{N_j,\Delta_j}$.
This is equivalent to the following condition:
\begin{align}
    |\Delta_j-\Delta_k|>\frac{N_j+N_k}{2} \quad \text{for} \quad 1\le \forall j< \forall k \le n.
\end{align}
We define a connected type II excitation $\{I\}_{\bm{h}}$ as follows:
\begin{align}
    \{ I \}_{\bm{h}}:=\bigcup_{j=1}^n \{I\}_{N_j,h_j,\Delta_j}.
    \label{eq:connected-typeII}
\end{align}
A connected type II excitation $\{I\}_{\bm{h}}$ is specified by the positions of holes $\bm{h}=(h_1,h_2,\cdots,h_n) \in H$, where we introduce a set of hole positions as the direct product $H=\prod_{j=1}^n \{0,1,\cdots,N_j-1\}$.
We define a family of sets of connected type II excitations as
\begin{align}
    \Imath_{\text{C}}:=\{\{I\}_{\bm{h}} | \bm{h} \in H\}.
\end{align}
$\Imath_{\text{C}}$ has $C=\prod_{j=1}^n N_j$ elements.

\begin{figure}[bth]
    \centering
    \includegraphics[height=5.0cm]{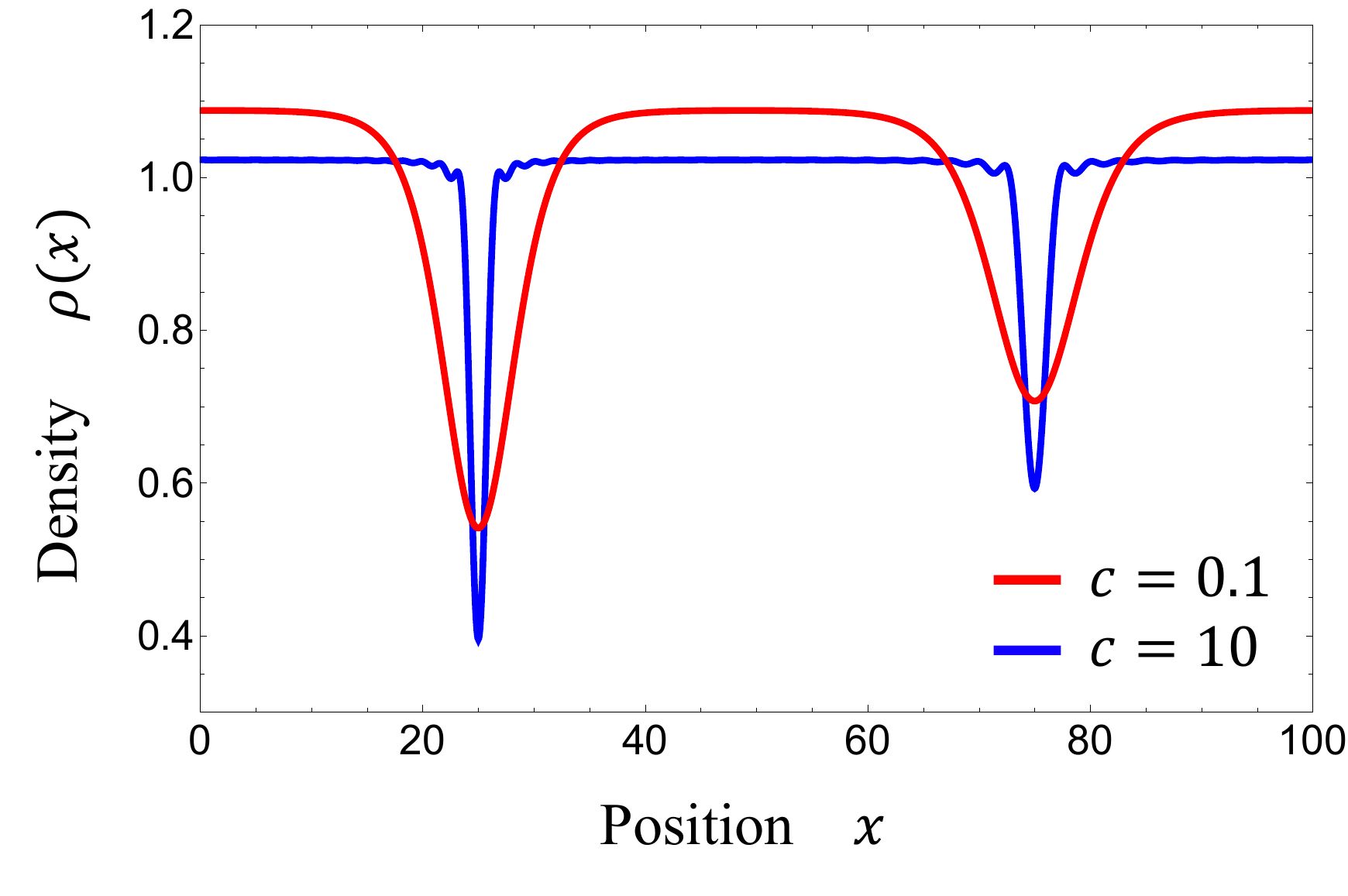}\\
    (a) Initial density profiles of the 2-quamtum soliton states \\
    \vspace{0.3cm}
    \includegraphics[height=5.0cm]{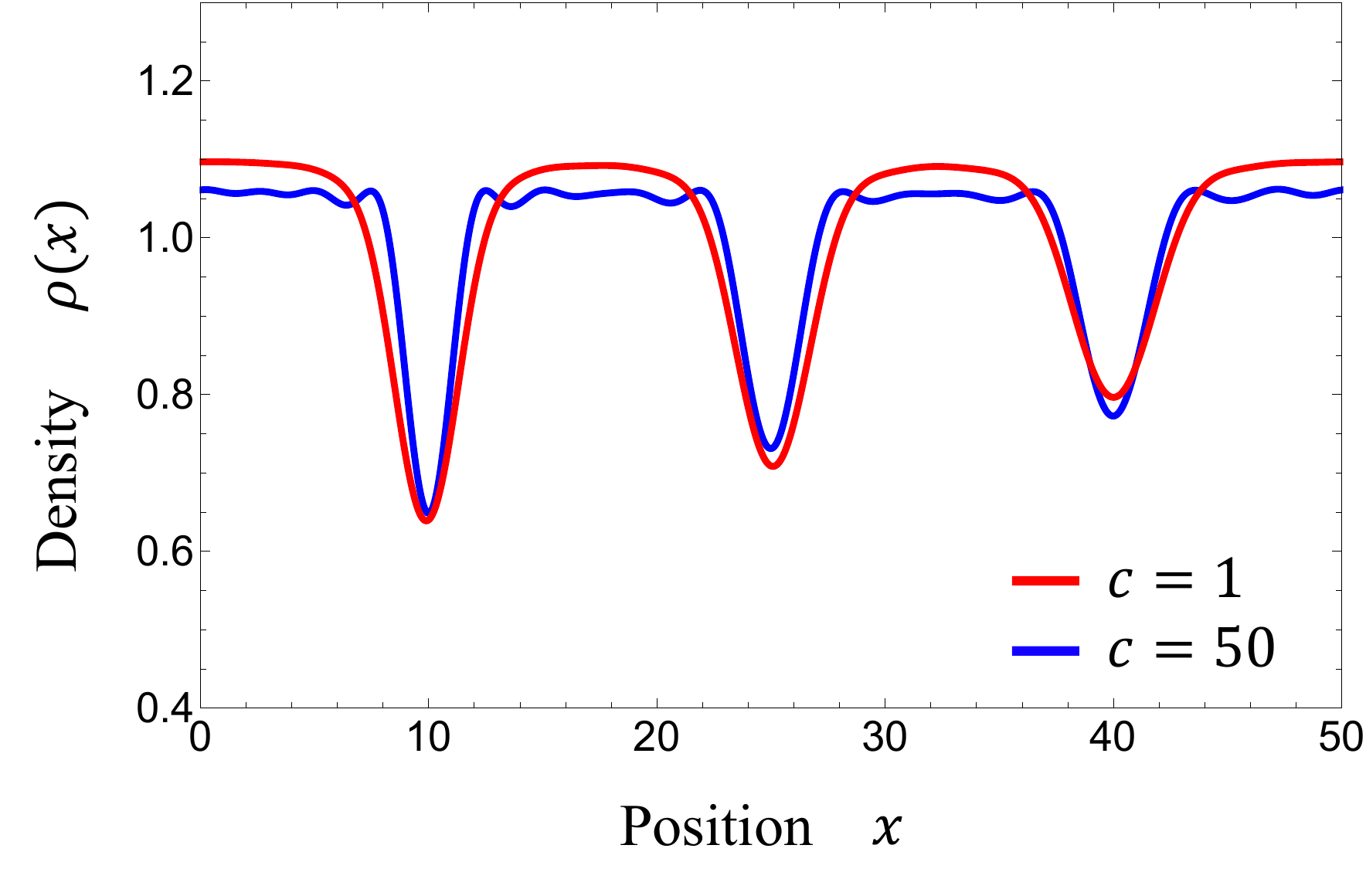}\\
        (b)  Initial density profiles of the 3-quamtum soliton states  \\
    \caption{Initial density profiles of the multi-quantum soliton states. 
    (a) Initial density profiles of the 2-quantum soliton states. Parameters are set to $L=N=100$, $N_1=60$, $N_2=40$, $\Delta_1=80$, $\Delta_2=-70$, $X_1=\frac{L}{4}$, and $X_2=\frac{3L}{4}$. The coupling constants are $c=0.1$ (red line) and $c=10$ (blue line).
    (a) Initial density profile of the 3-quantum soliton states. Parameters are set to $L=N=50$, $N_1=20$, $N_2=16$, $N_3=14$, $\Delta_1=40$, $\Delta_2=10$, $\Delta_3=-50$, $X_1=\frac{L}{5}$, $X_2=\frac{L}{2}$, and $X_3=\frac{4L}{5}$. The coupling constants are $c=1$ (red line) and $c=50$ (blue line).
    Multi-dark solitons are observed in the initial density profiles. 
    The depth of the $j$th soliton depends on the particle number $N_j$, and it increases as $N_j$ increases. 
    The shapes of the solitons also depend on the coupling constant $c$, and it becomes sharper as $c$ increases.
    }
    \label{fig:initial-profile}
\end{figure}

An $n$-quantum soliton state is constructed by superposing all eigenstates in $\Imath_{\text{C}}$.
The total momentum of $\{ I \}_{\bm{h}} \in \Imath_{\text{C}}$ is given by  $P_{\bm{h}}=\sum_{j=1}^n P_{h_j}$, where $P_{h_j}=\frac{2\pi (h_j+N_j\Delta_j)}{L}$ ($0 \le h_j \le N_j-1$).
We denote the normalized Bethe eigenstate corresponding to the connected type II excitation $\{ I \}_{\bm{h}}$ as $|P_{h_1},P_{h_2},\dots,P_{h_n}\ket$. 
We introduce an $n$-quantum soliton state $|X_1,X_2,\cdots,X_n\ket$ as follows:
\begin{align}
\begin{split}
    |X_1,&X_2,\cdots,X_n\ket \\
    =&\frac{1}{\sqrt{C}} \sum_{\bm{h}\in H} 
    \text{exp} \left[ -i \sum_{j=1}^n P_{h_j} Q_j \right] |P_{h_1},P_{h_2},\dots,P_{h_n}\ket,
\end{split}
\label{eq:multi-soliton}
\end{align}
where $Q_j$'s are parameters that determine the position of the $j$th soliton as $X_j=Q_j+\frac{L}{2}$.
As we show later, $n$-dark solitons emerge in the density profile of this state (Fig. \ref{fig:initial-profile}). Each $\Imath_{N_j,\Delta_j}$ corresponds to the $j$th soliton.

Next, we consider the time evolution of the density profile of the $n$-quantum soliton states.
We denote the $n$-quantum soliton state at time $t$ by $|X_1,X_2,\cdots,X_n;t\ket=\text{exp}(-i\HLL t)|X_1,X_2,\cdots,X_n\ket$.
The expectation value of the density operator $\hat{\rho}(x)=\pd(x)\ps(x)$ in the $n$-quantum soliton state at time $t$ is given by
\begin{align}
\begin{split}
    \bra X_1,&X_2,\cdots,X_n;t| \hat{\rho}(x) | X_1,X_2,\cdots,X_n;t \ket \\
    =&\frac{1}{C} 
    \sum_{\bm{h},\bm{h'}\in H} 
    \text{exp}\left[ i \sum_{j=1}^n (P_{h'_j}-P_{h_j})Q_j \right.\\
    &\left. - i(P_{\bm{h'}}-P_{\bm{h}})x + i(E_{\bm{h'}}-E_{\bm{h}})t
    \right] \\
    &\hspace{1.0cm} \bra P_{h'_1},P_{h'_2},\dots,P_{h'_n}|\hat{\rho}(0)|P_{h_1},P_{h_2},\dots,P_{h_n}\ket,
\end{split}
\label{eq:time-evo-density}
\end{align}
where $E_{\bm{h}}$ denotes the energy eigenvalue of the connected type II excitation  $|P_{h_1},P_{h_2},\dots,P_{h_n}\ket$. 
Using the determinant formulas based on the Bethe ansatz technique, the form factors $\bra P_{h'_1},P_{h'_2},\dots,P_{h'_n}|\hat{\rho}(0)|P_{h_1},P_{h_2},\dots,P_{h_n}\ket$ can be evaluated efficiently \cite{gaudin1983fonction,korepin1982calculation,slavnov1989calculation,slavnov1990nonequal,Calabrese_2007}.

Based on Eq. (\ref{eq:time-evo-density}), we numerically calculated the time evolution of the density profiles of the multi-quantum soliton states.
Fig. \ref{fig:initial-profile} shows the initial density profiles ($t=0$) of (a) the 2-quantum soliton states and (b) the 3-quantum soliton states.
Two and three localized solitary waves corresponding to dark solitons are observed in the density profiles.
The shape of the $j$th soliton depends on the particle number $N_j$ of the $j$th component $\Imath_{N_j,\Delta_j}$.
The depth of the $j$th soliton increases as the particle number $N_j$ increases.
In the simulations, we varied the coupling constant $c$.
As in the case of single-quantum solitons, the shape of the solitons also depends on the value of $c$. It becomes sharper as $c$ increases.

\begin{figure}[tbh]
    \centering
    \includegraphics[height=13.0cm]{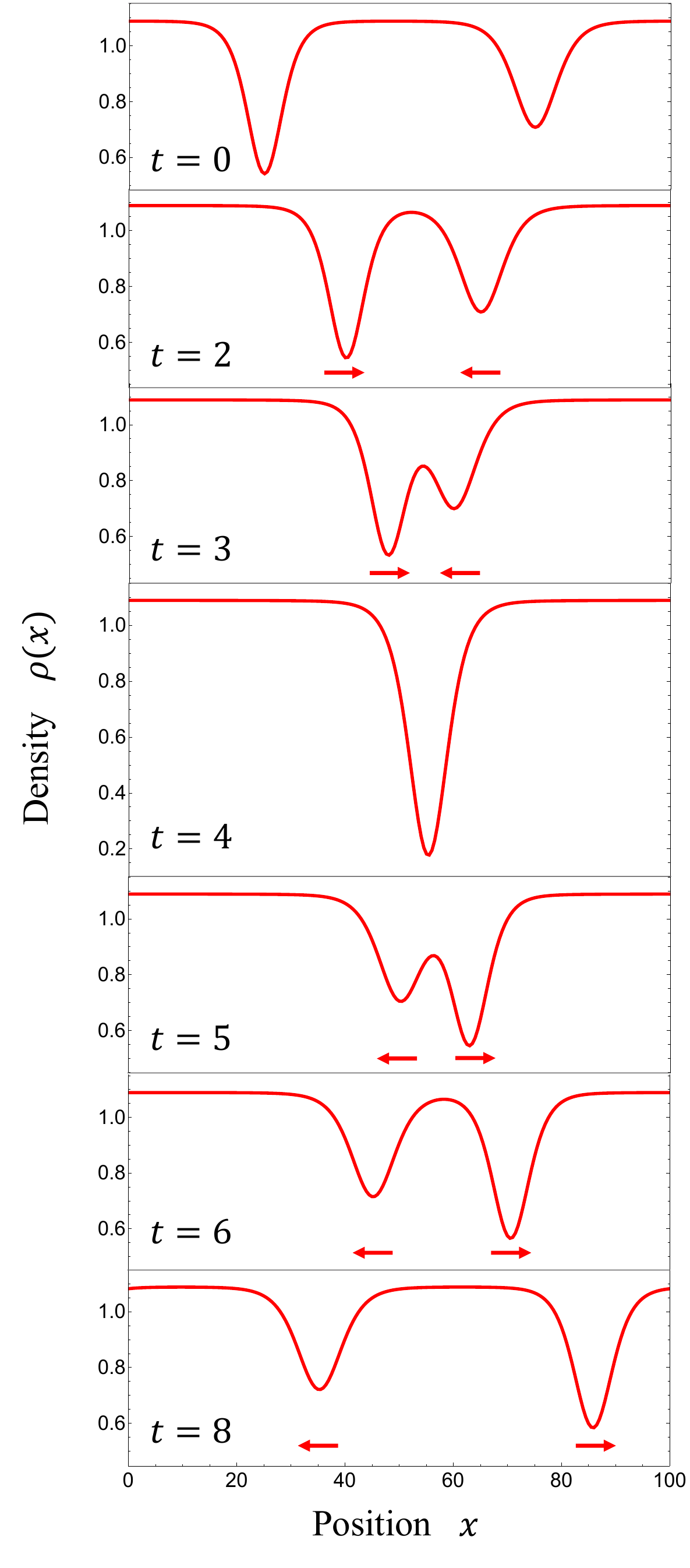}
    \caption{Time evolution of the density profile of the 2-quantum soliton state. Parameters are set to $L=N=100$, $N_1=60$, $N_2=40$, $\Delta_1=80$, $\Delta_2=-70$, $X_1=\frac{L}{4}$, $X_2=\frac{3L}{4}$, and $c=0.1$. The scattering of the quantum-dark solitons is observed from a series of snapshots. Each localized solitary wave behaves like a particle, and is stable against the collisions.}
    \label{fig:dynamics}
\end{figure}

Fig. \ref{fig:dynamics} shows the time evolution of the density profile of the 2-quantum soliton state. 
Each localized solitary wave behaves like a particle. 
The scattering of quantum dark solitons is observed via simulation.
At the early time, two localized solitary waves move towards each other. 
The velocity of the $j$th soliton depends on the $\Delta_j$ of the $j$th component $\Imath_{N_j,\Delta_j}$.
Then, they collide. Subsequently, they pass through each other and continue moving without changing their shapes and velocities.
As in the case of single-quantum solitons, wave packets of multi-quantum solitons gradually disperse due to quantum effects.
However, quantum solitons retain the properties of classical solitons: stability against collisions.

In the following, we consider the interaction of quantum dark solitons.
It is well known that, in the case of the scattering of classical solitons, the positions of the solitons are shifted.
Do such position shifts occur in the case of quantum solitons?
We observed the position shifts of the quantum dark solitons in the weak coupling regime.

\begin{figure}[tbh]
    \centering
    \includegraphics[height=5.0cm]{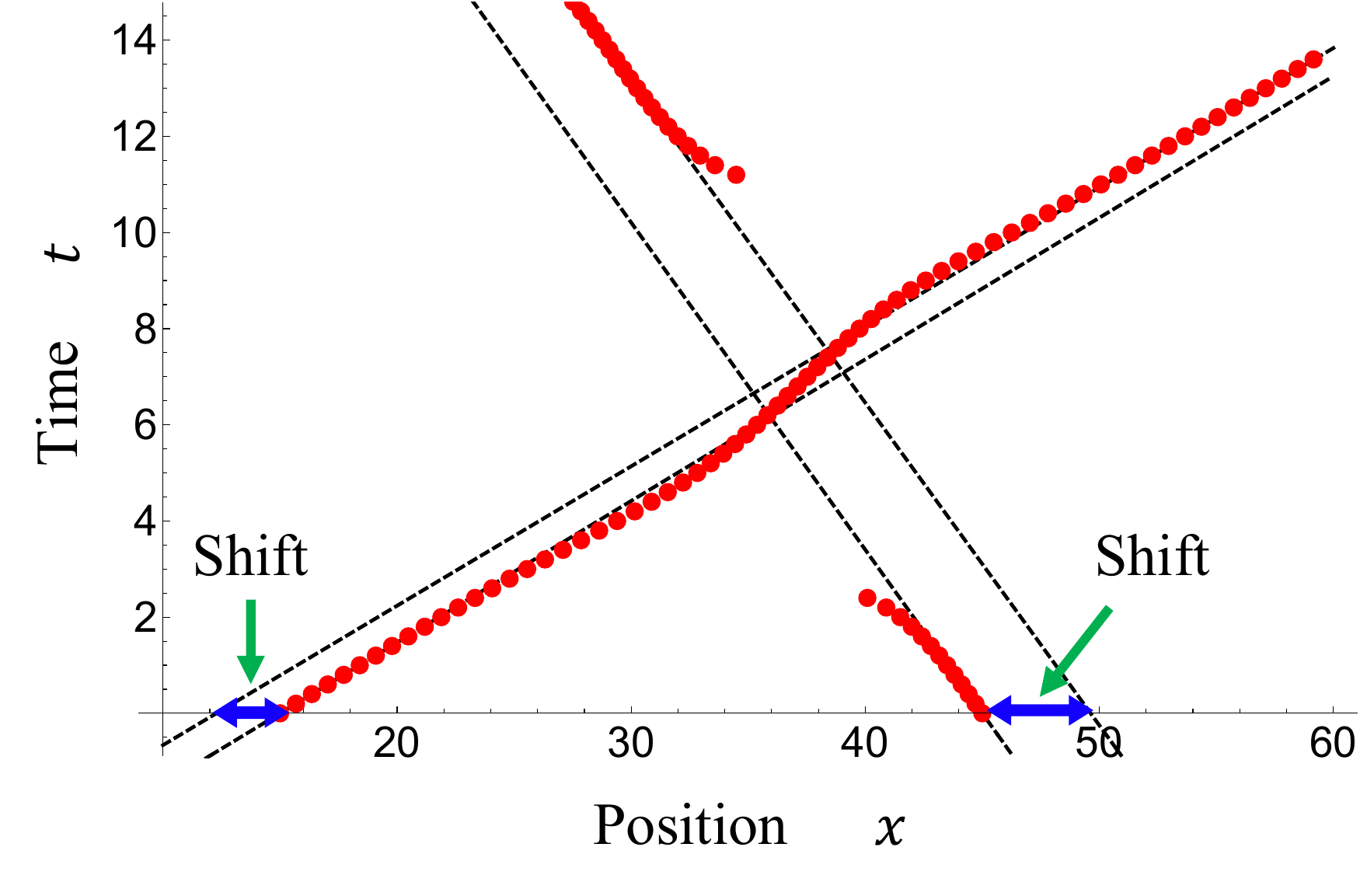}
    \caption{Trajectories of the quantum solitons during the collision. 
    Parameters are set to $L=N=60$, $N_1=20$, $N_2=40$, $c=0.015$, $\Delta_1=-26$, $\Delta_2=25$, $X_1=\frac{3L}{4}$, and $X_2=\frac{L}{4}$. 
    The red dots show the positions of the solitons. The gray dotted lines show the hypothetical trajectories of the solitons when no scattering occurs. Position shifts are observed from the shift of the gray dotted lines.}
    \label{fig:scattering}
\end{figure}

\begin{figure}[tbh]
    \centering
    \includegraphics[height=5.0cm]{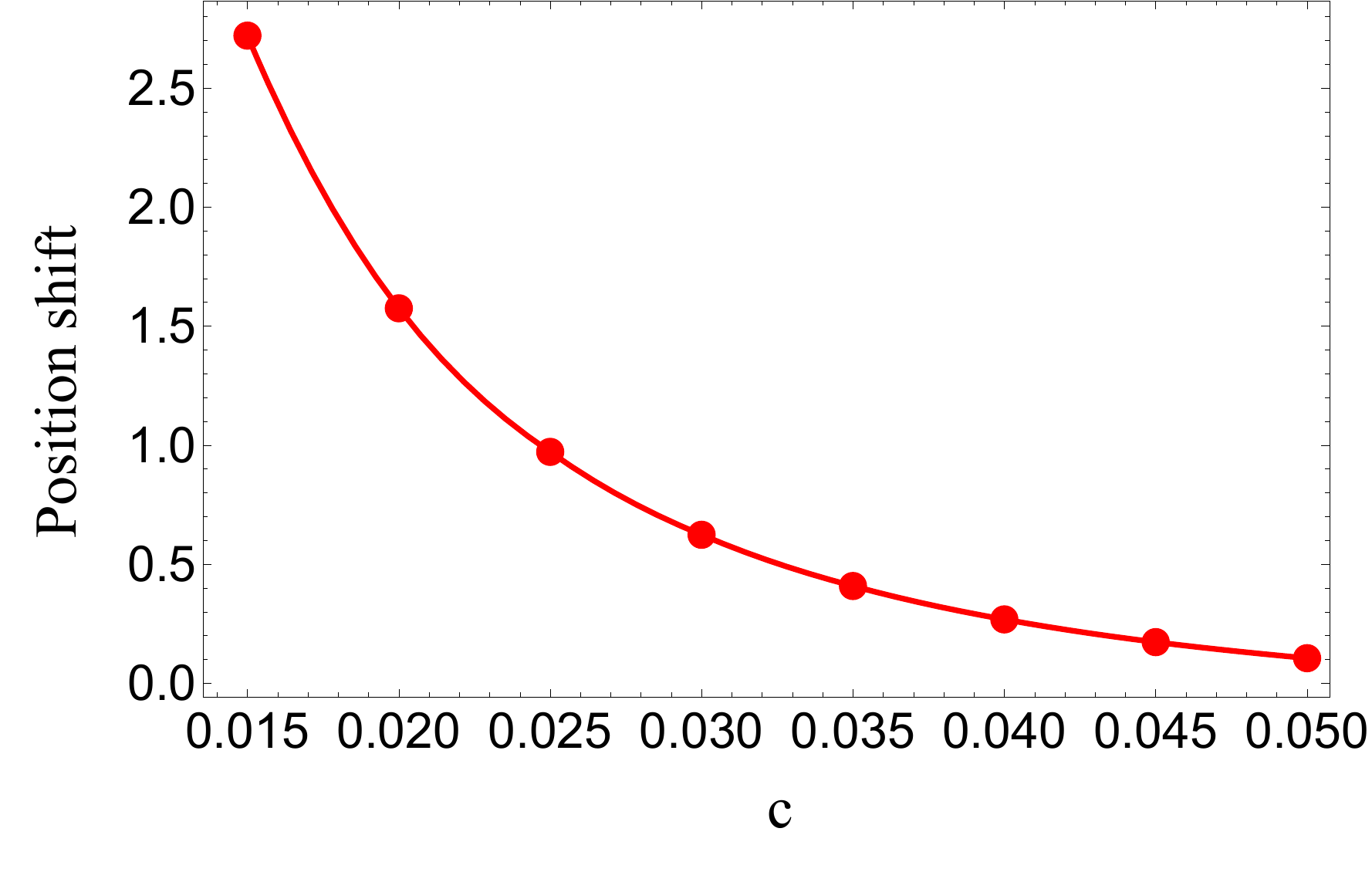}
    \caption{Dependence of the position shifts of the quantum solitons on the coupling constant $c$. Parameters are set to $L=N=60$, $N_1=20$,  $N_2=40$, $\Delta_1=-26$, $\Delta_2=25$, $X_1=\frac{3L}{4}$, and $X_2=\frac{L}{4}$.
    The red dots show the position shifts of the quantum soliton corresponding to $\Imath_{N_2=40,\Delta_2=25}$.
    The position shifts are observed in the weak coupling regime, and they decrease as $c$ increases.}
    \label{fig:position-shift}
\end{figure}

In Fig. \ref{fig:scattering}, we show the trajectories of the quantum solitons during the collision.
In the simulation, we consider the 2-quantum soliton state consisting of $\Imath_{N_1=20,\Delta_1=-26}$ and $\Imath_{N_2=40,\Delta_2=25}$.
The red dots show the positions of the solitons.
The gray dotted lines show the hypothetical trajectories of the solitons when no scattering occurs.
The collision occurs at around $t=7$.
Before and after the collision, the gray lines shifted parallel to the $x$ axis.
This implies that the positions of the solitons are shifted due to the collision.
The magnitude of the position shifts depends on the coupling constant $c$.
Fig. \ref{fig:position-shift} shows the dependence of the position shifts of the quantum dark solitons on the coupling constant $c$.
We consider the 2-quantum soliton state consisting of $\Imath_{N_1=20,\Delta_1=-26}$ and $\Imath_{N_2=40,\Delta_2=25}$, and show the position shifts of the soliton corresponding to $\Imath_{N_2=40,\Delta_2=25}$ in Fig. \ref{fig:position-shift}.
As shown in Fig. \ref{fig:position-shift}, the position shifts are observed in the weak coupling regime, and they decrease as $c$ increases.
This result is consistent with the fact that the weak coupling limit corresponds to the classical limit, and the quantum dark soliton approaches the classical one in this limit.

\textit{Conclusion.}---
In this Letter, we established the construction of multi-quantum dark soliton states and elucidated the nature of their scattering in the quantum integrable system. 
Through numerical calculations based on the Bethe ansatz, we showed that multi-dark solitons emerged in the density profile of the multi-quantum dark soliton states.
In addition, we realized the scattering of quantum dark solitons, and revealed that quantum dark solitons retain the fundamental properties of classical solitons: stability against collisions and position shifts due to interactions.
These are the first results to show the correspondence between quantum states constructed from type II excitations and classical dark solitons in view of interactions among multi-solitons.
In the present work, we consider the construction of multi-quantum solitons in the LL model.
However, classical solitons universally appear in classical systems, and
quantum solitons are also expected to appear in a variety of quantum integrable systems.
In future works, we will generalize the construction of multi-quantum solitons in the LL model for general quantum integrable systems, and develop a  theoretical formulation of quantum solitons.

\begin{acknowledgments}
The authors thank Tetsuo Deguchi for useful comments.
This work was supported by JSPS KAKENHI Grant Number JP18K03448.
\end{acknowledgments}





\nocite{*}

\bibliography{apssamp}

\end{document}